# Temperature response of Earth to the annual solar irradiance cycle

David H. Douglass,[a] Eric G. Blackman,[a,b] and Robert S. Knox[a,b]*

[a]Department of Physics and Astronomy, University of Rochester, Rochester, NY 14627–0171
[b]Laboratory for Laser Energetics, University of Rochester, Rochester, NY 14623–1299

## Abstract

We directly determine the sensitivity and time delay of Earth's surface temperature response to annual solar irradiance variations from 60 years of data. A two-layer energy balance model is developed to interpret the results. Explaining both the resulting low sensitivity and time delay of 1-2 months requires negative feedback.

## PACS codes and keywords
92.70.Gt    Climate dynamics
91.10.Vr    Ocean/Earth/atmosphere interaction
96.40.Kk    Solar modulation and geophysical effects
96.35.Hv    Neutral atmospheres

*Corresponding author.
    Dept. of Physics and Astronomy, Univ. of Rochester, Rochester, NY 14627-0171 USA
    Phone 1-585-275-8572
    Fax 1-585-273-3237
    Email `rsk@pas.rochester.edu`

# 1. Introduction

The average surface temperature, $T$, of terrestrial planets such as Earth depends on a variety of factors [1], and distinguishing solar and geophysical influences from those of greenhouse gases is of great current interest. A particularly important and straightforward effect is the response of $T$ to variations $\Delta I$ in the solar irradiance $I$, characterized by the amplitude of temperature variation $\Delta T$ and a phase delay $\phi$. The seasonal cycle is one of the largest climate changes observed. It is a common experience at mid-latitudes to note the large seasonal variation of the temperature, of order tens of degrees, and its phase lag of 1 to 1.5 months behind the corresponding large $\Delta I$ in the solar irradiance. The scientific task is to explain both the observed sensitivity $k$ (the ratio of $\Delta T$ to $\Delta I$), and the time delay. A simple no-feedback radiation model fails because it predicts a $\Delta T$ much larger than observed for other than global averages.

We wish to compare seasonal data with an energy balance model (EBM) in order to understand the observed sensitivity. While the global average temperature has a very small annual component resulting from the ellipticity of Earth's orbit [2], the usual seasonal effect at specific latitudes provides a much larger signal. It can be brought out clearly by averaging mean monthly temperatures over a large data set, in our case covering 60 years. Spatial averages of these data over rather wide latitude zones enable comparison with EBM predictions.

EBMs featuring surface temperature as the dependent variable have been studied in detail by North and colleagues [3–5]. These models are based on an empirical relationship [6] between the outgoing infrared radiation at the top of the atmosphere and the surface temperature.  For global averages, these particular models predict that radiative fluxes must be accompanied by positive feedback.  Although they provide a qualitative picture and many verisimilar quantitative predictions, certain serious discrepancies exist, particularly in the prediction of phase delays larger than observed in the extratropical latitudes (see [4], Figs. 3a and 3b).  Our determined zonal averages confirm the existence of the shorter phase lags and produce zonal values of the sensitivity of the solar forcing that indicate a need for  large negative feedback over that part of the globe studied (60S-60N).  These findings are consistent with an extended EBM that employs two layers and includes both the greenhouse effect and feedback explicitly.  This model has a quasi-one-layer limit that can be compared with the North EBM equation.

## 2.  Data and analysis

The average solar irradiance at given latitudes and longitudes can be determined from the solar constant by standard methods [3,7].  For temperature, we use the most complete set of seasonal data, which have been compiled and interpolated to a grid composed of 0.5° x 0.5° cells from ten sources spanning the years 1920 to 1980 [8].  They give the 12 monthly averages for each cell.  At each latitude we average the cell values over longitudes, then over four 30-degree zones (0-30 N, S and 30-60 N, S) as shown in Table 1.  The computed irradiance, including a correction for the ellipticity of Earth's orbit, is similarly treated.

Figure 1 shows *I* and *T vs*. month for the zones described above.  Comparison of Figs. 1(a) and 1(b) shows a striking correlation between them with $T$ lagging $I$ in phase by 1-2 months. The variation of $I$ and $T$ around their averages can be interpreted empirically to sufficient accuracy by the equations

$$\Delta I = \Delta I_0 \cos \omega t, \quad \Delta T = A \cos(\omega t - \phi), \tag{1}$$

where $\omega$ is the forcing frequency ($2\pi$/year) and where $A$ and $\phi$ are determined by making phase plots of $T$ (ordinate) vs. $I$ (abscissa). Because of the periodicity, these plots are closed Lissajous curves. If $T$ lags (leads) $I$ by a phase angle $\phi$, the loop's area is determined by $\phi$ and the phase point moves counter-clockwise (clockwise). We have plotted $T(t)$ vs. $I(t - t_d)$, where $t_d$ is an imposed phase delay varied in increments of 0.5 months. [The phase shift in radians is expressed in months via $\phi(\text{rad}) = 2\pi t_d(\text{mo})/12$.] The best fit value of $\phi$ is determined by finding the $t_d$ that yields the loop of smallest area. The technique determines both $\phi$ and $A$, the latter from the slope of the $T$ vs. $I$ line.

Figure 2 shows the phase plots of $I$ and $T$ for the various zones. In the southern temperate zone (30-60 S, Fig. 2a), the plots are for phase delays of $t_d =$ 1.0, 1.5, and 2.0 months. The 1.0-mo delay plot shows a counterclockwise trajectory indicating that the delayed $T$ lags $I$. The plot for $t_d = 2.0$ mo shows the opposite. The plot for $t_d = 1.5$ mo gives the minimum area and yields $\phi$. For this curve, a line has been least-squares determined giving a regression coefficient of $k_m = A/\Delta I_0 = 0.015$ K/(W/m$^2$) with correlation coefficient $R^2 = 0.981$. From peak-to-peak differences, we obtain $T_{pp}/I_{pp} = \Delta T/\Delta I = 5.1/350 = 0.015$ K/(W/m$^2$), the same as the regression coefficient $k_m$. The regression method has the advantage that the uncertainties in $k_m$ and $\phi$ can be minimized by maximizing $R^2$. The values of $\phi$ listed in Table 1 were found by plotting $R^2$ vs. $t_d$ and choosing that $t_d$ which maximized $R^2$. The results for the southern tropic zone, Fig. 3B, are similar to those of the southern temperate zone. Here the best fit parameters were $k_m = 0.020$ and $\phi = 1.5$ mo. These values are also listed in Table 1, along with the results from the northern tropic (0-30 N, Fig. 2c), with best fit values $k_m = 0.022$ and $\phi = 1.0$ mo, and the northern temperate (30-60 N, Fig. 2d) with best fit values $k_m = 0.055$ and $\phi = 1.0$ mo.

## 3. Global energy balance model

To interpret the data we employ a simple EBM that treats the surface and atmosphere layers separately. The atmosphere layer is frequently introduced as a "black shield" to explain the greenhouse effect semiquantitatively (*e. g.,* [3]). We consider the atmosphere layer to be an essential part of our EBM and assign to it an absorptivity and emissivity $\varepsilon$. Further refinements of this two-layer Arrhenius model involving atmospheric reflectivities and absorptivities and non-radiative fluxes are available [9,10] but we will not make use of them here. As in other EBMs, a fraction $\alpha$ of the solar irradiance is reflected by Earth's atmosphere and the remainder heats the surface below. The surface re-radiates the incident energy, mostly in the infrared. A fraction $\varepsilon$ is absorbed by the atmosphere and the rest passes through into space. The atmosphere then radiates part to Earth and part to space. The irradiance at the surface is then the sum of that coming directly from the sun, plus that which is radiated downward by the atmosphere. The respective radiative energy flux balance equations for the atmosphere and surface layers in steady state are

$$2F_A - \varepsilon F_S = 0 \qquad\qquad (2a)$$

$$F_S - F_A = F_i,  \tag{2b}$$

where $F_A = \varepsilon\sigma T_A{}^4$ and $F_S = \sigma T^4$ are respectively the flux radiated from each surface of the atmosphere layer (temperature $T_A$) and from the surface layer (temperature $T$); $\sigma$ is the Stefan-Boltzmann constant and $F_i$ is the solar flux incident at the surface. When the model is applied globally and time averaged, $F_i = (1 - \alpha)I_0$, where $I_0 = 342$ W/m$^2$ is the solar constant. Solution of these equations for the surface flux produces in this case a global average surface temperature

$$T_0 = \left[\frac{1-\alpha}{\sigma(1-\varepsilon/2)}I_0\right]^{\frac{1}{4}}.  \tag{3}$$

(Note – a "4" has been removed from original Eq. 3.) To reproduce Earth's average temperature of 288 K, the values $\alpha = 0.33$ and $\varepsilon = 0.83$ may be used.

To consider the time variations in the fluxes around their mean values, we express all quantities in terms of these variations and subtract the time-averaged atmosphere energy equation, obtaining for the atmosphere and the surface, respectively,

$$c_A d\Delta T_A/dt + 2\Delta F_A - (\varepsilon - f_A)\Delta F_S = 0,  \tag{4a}$$

$$c_S d\Delta T/dt - \Delta F_A + (1 - f_S)\Delta F_S = \Delta F_i,  \tag{4b}$$

where $\Delta$ represents the difference between a quantity measured at *t* and its mean, $c_A$ is the effective heat capacity of the atmosphere per unit projected area and $c_S$ is the effective heat capacity per unit area of the thermal layer (dominated by the oceans). $f_A$ and $f_S$ are parameters that account for feedback effects on the atmosphere and surface, respectively, due to the dependence of model parameters on surface temperature. They contribute to an effective overall feedback, as discussed below (Eq. 7). The coupled equations (4) and their solutions will be referred to as the KBD model. By assuming that $F_S$ depends on time only through variations in $T$, Taylor expanding, and keeping linear terms, we have $\Delta F_S = q(T_0)\Delta T$ for $\Delta T << T_0$, where $T_0$ is the time-averaged surface temperature, and where $q(T_0) = 4\sigma T_0^3$, with a similar expansion of $\Delta F_A$. The coupled equations (4) then determine $\Delta T(t)$ and $\Delta T_A(t)$.

The full KBD model is developed in a subsequent paper [11]. However, for present purposes an informative bridge between this model and single-temperature-variable EBMs can be built. Rearranging (4a) as an expression for $\Delta F_A$ and substituting the result into (4b) results in

$$\frac{c_A}{2}\frac{d\Delta T_A}{dt} + c_S\frac{d\Delta T}{dt} + q(T_0)(1-\tilde{f})\Delta T = \Delta F_i, \qquad (5)$$

where $\tilde{f} = f_S - f_A/2 + \varepsilon/2$. Eq. (5) can be compared to a globally-averaged North-Budyko EBM equation [3,4], if we note that there is a dynamical contribution from the upper layer represented by the first term, and that an explicit form of the Budyko radiation damping coefficient is given: $B = q(T_0)(1 - \tilde{f})$. Here we consider two limiting cases of Eq. (5) that provide some immediate insight into the properties of the data.

*Case 1. Time-independent solution.* Consider a step-function increase of solar irradiance by an amount $\Delta I_0$. After transients have died away, Eq. (5) will be satisfied with all time derivatives set to zero. The shift in $T_0$ is therefore given by

$$\Delta T_0 = \frac{(1-\alpha)\Delta I_0}{q(T_0)(1-\tilde{f})} = \frac{1}{1-f_{\text{eff}}}\frac{T_0}{4}\frac{\Delta I_0}{I_0}, \qquad (6)$$

where $f_{\mathrm{eff}}$ is an effective feedback

$$f_{\mathrm{eff}} = \frac{f_S - \frac{1}{2} f_A}{1 - \frac{1}{2} \varepsilon}. \tag{7}$$

The sensitivity of the surface temperature to very slow variations is therefore

$$k = \frac{\Delta T_0}{\Delta I_0} = \frac{k_{NF}}{1 - f_{\mathrm{eff}}}. \tag{8}$$

The no-feedback sensitivity

$$k_{NF} = \frac{T_0}{4 I_0} \tag{9}$$

is $288/(4 \times 342) = 0.211$ K/(W/m$^2$) for global averages.

*Case 2.  $c_A$ very small compared with $c_S$.*  If the first term of Eq. (5) may be dropped, the equation may be written

$$\Delta T + \tau \frac{d \Delta T}{dt} = k \Delta I, \tag{10}$$

where $k$ is the same as in *Case 1* and we define a relaxation time

$$\tau = c_S / [q(T_0)(1 - \tilde{f})] = c_S / B. \tag{11}$$

When $\Delta I = \Delta I_0 \cos \omega t$, Eq. (10) has the solution

$$\Delta T(t) = \frac{k \Delta I_0}{\sqrt{1 + (\omega \tau)^2}} \cos(\omega t - \phi), \tag{12}$$

where

$$\phi = \tan^{-1}(\omega \tau). \tag{13}$$

This solution reduces to Eq. (8) at zero frequency and provides a physical interpretation of the parameters $A$ and $\phi$ found in our empirical data analysis.  In particular, comparing with Eq. (1), we find

$$A = \frac{k\Delta I_0}{\sqrt{1 + (\omega\tau)^2}} = \frac{k\Delta I_0}{\sqrt{1 + \tan^2\phi}} = k_m\Delta I_0, \tag{14}$$

where $k_m$ is the measured sensitivity given in Table 1.

In the KBD model the complete solution of Eqs. (4) for $\Delta T$ is shown to have the form (1) in which the phase and amplitude depend on all the model parameters including $c_A$. Given $c_A$, the effective depth of the ocean mixing layer that determines $c_S$ can be inferred from the solution of Eqs. (4), revealing that $c_A < c_S$ is consistent for most of Earth. This motivates the illustrative approach of solving only Eq. (10).

## 4. Zonal EBM

The EBM as thus far described applies strictly to certain whole-Earth solar flux variations of rather small magnitude. We apply the KBD model locally by assuming that during a typical time-averaging period the relevant energy fluxes are principally vertical. Spatial averages are taken over relatively wide latitude zones (see Table 1), eliminating the need to consider East-West transport except in the eventual interpretation of the effective heat capacities within a zone ([3], p. 100). Our double short-period time averaging (over 30 days and then over 60 years at corresponding times of the year) should minimize the effect of north-south transport, which is principally a seasonal phenomenon. Zonal averaging produces two kinds of quantitative effects as compared with whole-Earth averages. There is a larger annual variation in insolation because of the orbital effect and each zone has its own base or steady-state fluxes and temperatures that differ from $T_0 = 288$ K. Zonal parameters are presented in Table 1, identified by the angular brackets. For $\langle k_{NF} \rangle$, with sufficient accuracy simple averages of the maximum and minimum values of $T$ and $I$ were used for $\langle T \rangle$ and $\langle I \rangle$, respectively.

The determination of $k$ proceeds as follows: for each zone, from the measured phase $\phi$ and the known $\omega$, the value of $\tau$ is deduced from Eq. (13). From the measured peak-to-peak values the ratio $A/\Delta I_0$ is obtained, giving $k_m$. This value and that of $\phi$ allow us to deduce the zero-frequency sensitivity $k$ (Eq. 8). There is one potential limitation to the method: finding $k/k_m$ depends on knowing $\tan\phi$. In the event that $\phi \sim \pi/2$ (equivalent to 3 mo in our units), one knows only that $\tan\phi \gg 1$, and neither $\tau$ nor the ratio $k/k_m$ can be determined accurately. In our case, $\phi$ is of the order 1-2 mo and no problem is encountered. This limitation is a severe one in the EBM fitting cited earlier [4].

Now having determined $k$ and a value $k_{NF}$ for each zone, we are in a position to evaluate the effective feedback $f_{\text{eff}}$ using Eq. (8). As shown in Table 1, its values are large and negative in all of the zones studied. The parameter $f_{\text{eff}}$ is related to the "true" feedbacks $f_S$ and $f_A$ by Eq. (7) above. The denominator of Eq. (7) can be regarded as a renormalization of the feedbacks, increasing them, by the greenhouse effect. It appears here because of our consolidation of the two energy balance equations into one. In the full KBD treatment of the two equations the individual contributions of $f_S$ and $f_A$ can be considered. We do not attempt to do this here; our aim is to extract $k_m$ and $\phi$ from the data.

## 5. Discussion and conclusions

Fig. 3(a) shows the measured phase lags $\phi$ and corresponding $\tau$ obtained from (13). In the southern hemisphere $\phi \sim 1.5$ mo, and in the northern hemisphere $\phi \sim 1.0$ mo. This trend of decreasing $\phi$ is expected if $\tau$ is determined by $c_S$, since the south has a larger ratio of water (higher specific heat) to land (lower specific heat) than the north.

Fig. 3(b) shows a plot of $k_m$ and $k$ *vs.* latitude from the peak-peak difference values given in Table 1. Note that: (i) the correction due to the phase factor is not large, $\sim 10$ to 30%. (ii) The two tropical values have close values, suggesting no intrinsic north-south asymmetry. (iii) The southern extratropics has values roughly equal to tropical values.

To explain the observed $A$ in response to $\Delta I$ using Eqs. (10), $k$ and $\tau$ are needed. We have, for the first time, extracted both from data (Table 1). In the tropical region (30N to 30S), we find a seasonal delay (time constant) of $\tau = 1.02$ months, and a sensitivity of $k = 0.026$ K/(W/m$^2$), corresponding to $f_{\text{eff}} = -6.6$. The southern extratropics behaves like the tropics while the northern extratropics has a higher gain ($k$ closer to $k_{NF}$). The correlation with land/water ratio is apparent.

That we generally find $k \lesssim k_{NF}$ and thus $f_{\text{eff}} < 0$ for the 1/yr forcing frequency, differs from $f_{\text{eff}} > 0$ found for global irradiance variations associated with the 11-yr solar cycle [12]. These results are not contradictory because they apply to different forcing frequencies where different feedback processes might be presumed and, in particular, a zonal application of the theory differs from a global application.

Having the relevant relaxation times at hand, we can estimate the effective zonal heat capacities directly from the EBM relation $\tau = c_S/B = c_S k/(1 - \alpha)$, which follows from straightforward manipulation of Eqs. 6, 8, and 11. Zonal averages of the albedo, $\langle\alpha\rangle$, were evaluated from cellular values computed by Schmidt *et al.* [13], which in turn were based on estimates of surface and cloud reflectivities and correlated with satellite data ([14], p. 33). The values of $B$, $c_S$, and $\tau$ are listed in Table 1. These $\tau$ values are small compared to those of earlier studies, which assumed larger values as input constants. Our values suggest mixing layers of the order of 15-20 meters in the southern hemisphere. Such depths may well be consistent with the time scales involved in this analysis; the thermal diffusion length alone for one year is ~1.5 meters ([14], p. 85). Lukas [15] has determined that the mixing layer in the tropics is ~30 m.

The unusual nature of the small sensitivities discovered here can be seen in the very large values of the Budyko parameter $B$, which is often taken to be about 2 K/(W/m$^2$). As seen in Table 1, southern hemisphere values are 10 to 15 times this large, apparently a result of the large negative feedback. This may be related to recent findings that the radiation budget in the tropics has a high variability due to cloudiness changes [16].

White *et al.* [17] argue that the solar cycle may be coupled to the El Niño effect, possibly explaining $f_{eff} > 0$ for the solar cycle period. However, coupling to El Niño is weak on annual time scales so $f_{eff} < 0$ over these shorter periods is not inconsistent with this observation. Lindzen *et al.* [18] propose that negative feedback can arise if the atmosphere acts as an "infrared iris," although this has been challenged [19].

In summary, our results reveal most importantly that a simple EBM can accommodate both the small determined values of $A$ and the observed and commonly experienced, seasonal time delay of $\tau \sim$ 1-2 months. Previous applications of these models have been made in an effort to explain the seasonal temperature delay. However, they assumed from the beginning that $\tau$ was several years, leading to large $\omega\tau$ (and to phase lags $\phi \sim 3$ months) which allowed these models to explain the low measured sensitivities. But as we have shown the phase lags can be determined directly and are less than the 3 months assumed in these models. Therefore, the low sensitivities can be explained only by negative feedback that may represent cloud effects or convective transport between zones that does not average out. Nevertheless, the physics that determines the negative feedback remains an open question.

## Acknowledgements

DHD acknowledges support from the Rochester Area Community Foundation. The authors are indebted to Laura E. Schmidt for computing the monthly zonal insolation averages. EB thanks P. Goldreich for a related correspondence.

**Figure captions**

1. Zonal averages of insolation and surface temperature *vs*. month of the year.
   Tropic (0-30 N,S) and extratropic (30-60 N,S) zonal averages of (a) average insolation and (b) the 60-year and 30-day average of the observed surface temperature *vs*. month of the year. See text for sources.

2. Phase plots of temperature $T$ *vs*. insolation $I$ for the zones indicated. a. 60S to 30S; b. 30S to 0; c. 0 to 30N; d. 30 N to 60N

3. Latitude dependence of (a) phase $\phi$ and time constant $\tau$, (b) solar sensitivity. $k$(dynamic) corresponds to $k_m$ of the text, and $k$(zero frequency) refers to the zero-frequency sensitivity $k$ deduced from Eq. (8). For the sensitivity without feedback, see Table I and the discussion of $k_{NF}$ in the text. Phase plots of the data were also done in a small latitude zone at the equator, with the result $\tau = 1.1$ mo, $\phi = 1.0$ mo, $k = 0.029$, $k_m = 0.026$. The feedback estimated for this zone is $-5.0$.

**Table 1.** Measured and derived quantities for the latitude zones analyzed. The 3% variation of $I_0$ due to the ellipticity of Earth's orbit has been incorporated.

| latitude band → | 60S–30S | 30S–0 | 0–30N | 30N–60N |
|---|---|---|---|---|
| avg. latitude (deg) | –43.6 | –14.0 | 14.0 | 43.6 |
| $I_{max}$ (Wm$^{-2}$) | 509 | 476 | 448 | 479 |
| $I_{min}$ (Wm$^{-2}$) | 139 | 281 | 299 | 147 |
| $I_{pp} = I_{max} - I_{min}$ | 370 | 195 | 149 | 332 |
| $T_{max}$ (K) | 286.6 | 298.8 | 300.1 | 292.3 |
| $T_{min}$ (K) | 281.5 | 295.3 | 296.2 | 272.5 |
| $T_{pp} = T_{max} - T_{min}$ | 5.10 | 3.53 | 3.90 | 19.80 |
| $k_m = T_{pp}/I_{pp}$  K(Wm$^{-2}$)$^{-1}$ | 0.014 | 0.018 | 0.026 | 0.059 |
| $k_m$ = reg. coef. | 0.014 | 0.018 | 0.024 | 0.058 |
| phase $\phi$ (mo) | 1.52 | 1.48 | 0.94 | 1.10 |
| tan $\phi = \omega\tau$ | 1.02 | 0.98 | 0.54 | 0.65 |
| $\tau$ (mo) | 1.95 | 1.87 | 1.02 | 1.24 |
| $p = (1+\tan^2\phi)^{1/2}$ | 1.43 | 1.40 | 1.13 | 1.19 |
| $k = p\,k_m$   K(Wm$^{-2}$)$^{-1}$ | 0.020 | 0.025 | 0.027 | 0.069 |
| **Derived and related quantities:** | | | | |
| $\langle k_{NF}\rangle = \langle T\rangle/4\langle I\rangle$  (see text) | 0.219 | 0.196 | 0.200 | 0.226 |
| $\langle f_{eff}\rangle$  (from Eq. 8) | –10.1 | –6.8 | –6.4 | –2.2 |
| $\langle\alpha\rangle$  (see text) | 0.32 | 0.21 | 0.23 | 0.32 |
| $c_S = \tau(1-\langle\alpha\rangle)/k$ ($10^7$ Jm$^{-2}$K$^{-1}$) | 17.7 | 15.3 | 7.7 | 3.1 |
| $B = (1-\langle\alpha\rangle)/k$  (Wm$^{-2}$K$^{-1}$) | 34.6 | 31.2 | 28.5 | 9.6 |

Douglass-Blackman-Knox          FIGURE 1

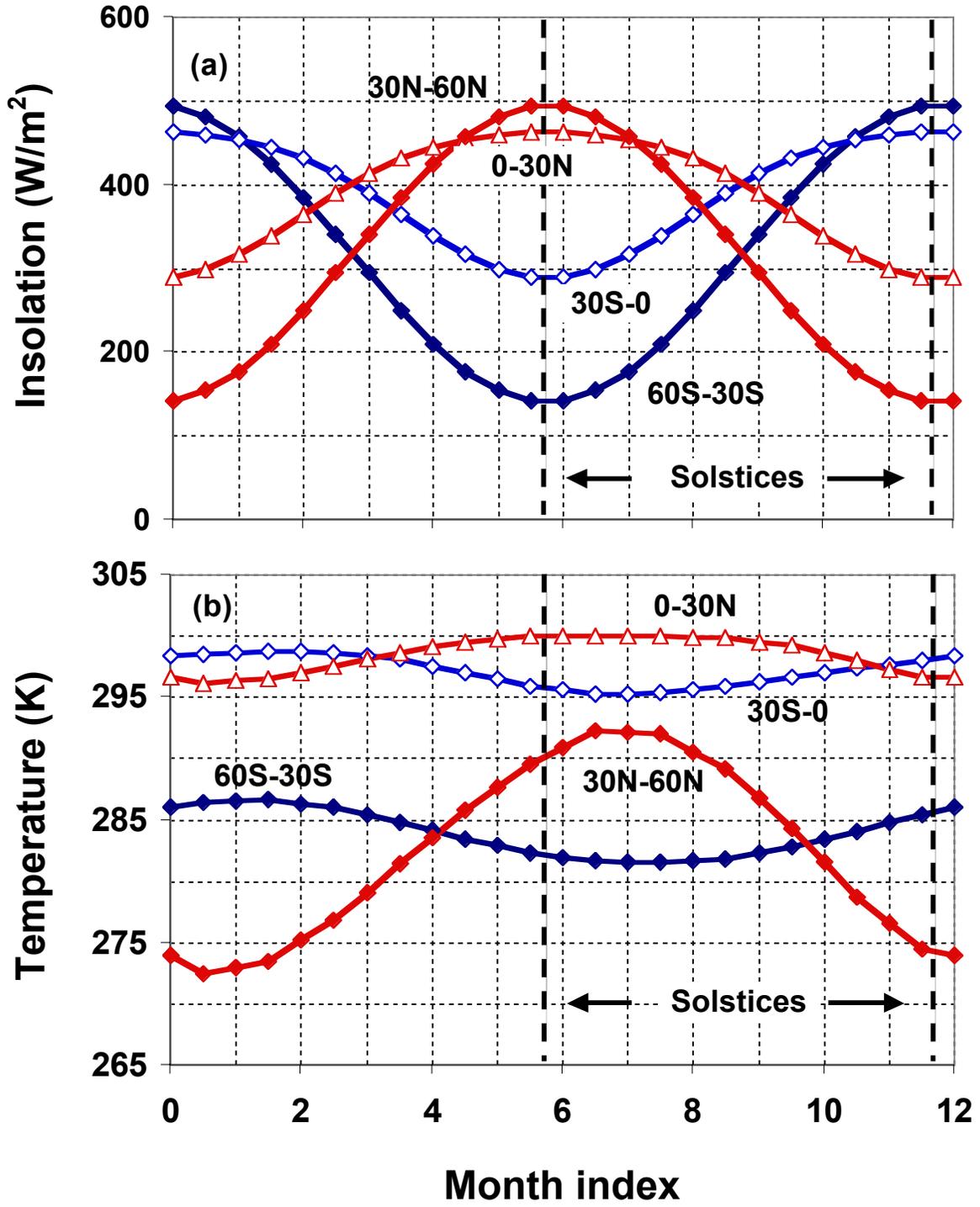

Douglass-Blackman-Knox    FIGURE 2

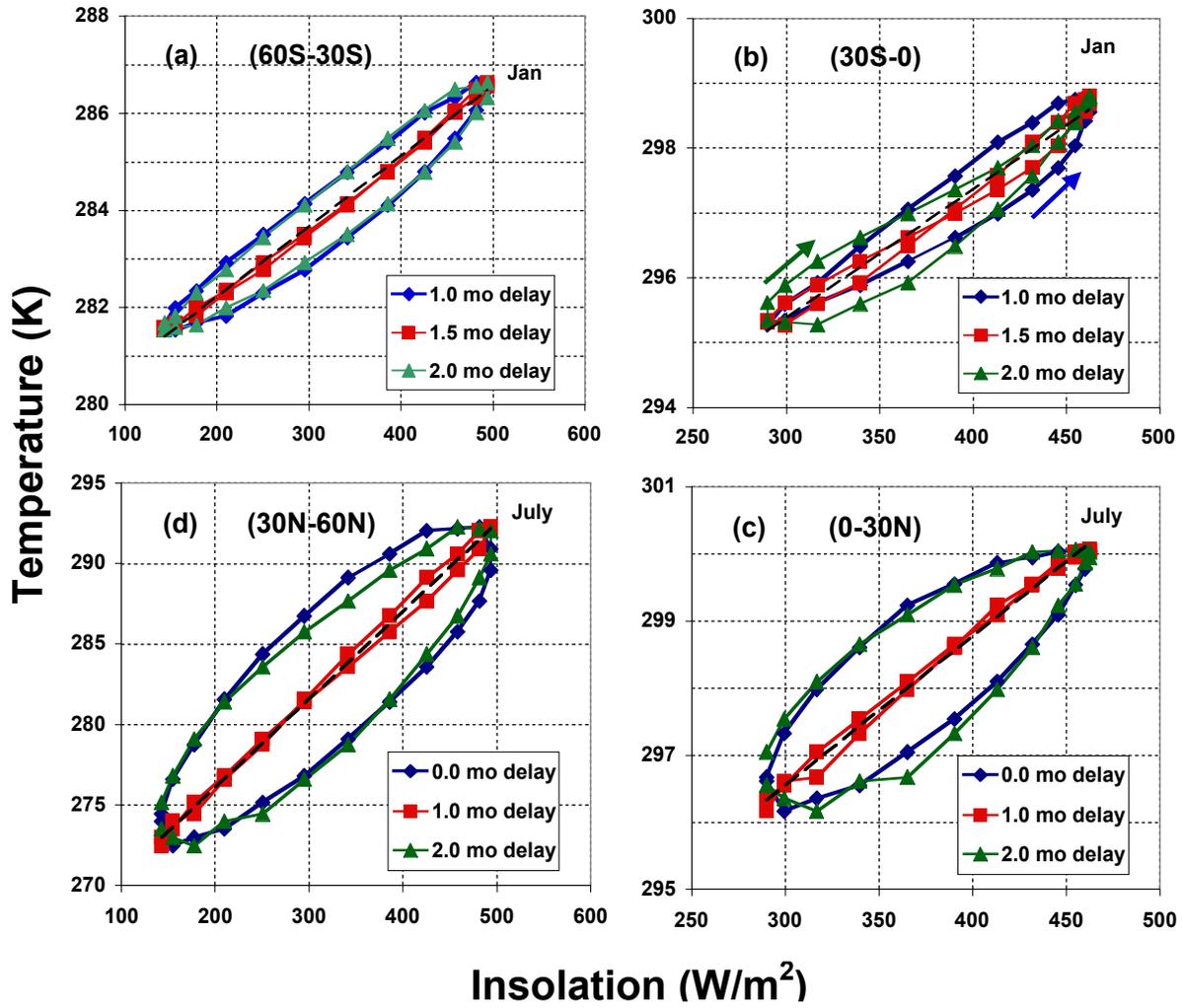

Douglass-Blackman-Knox (rev.)   FIGURE 3

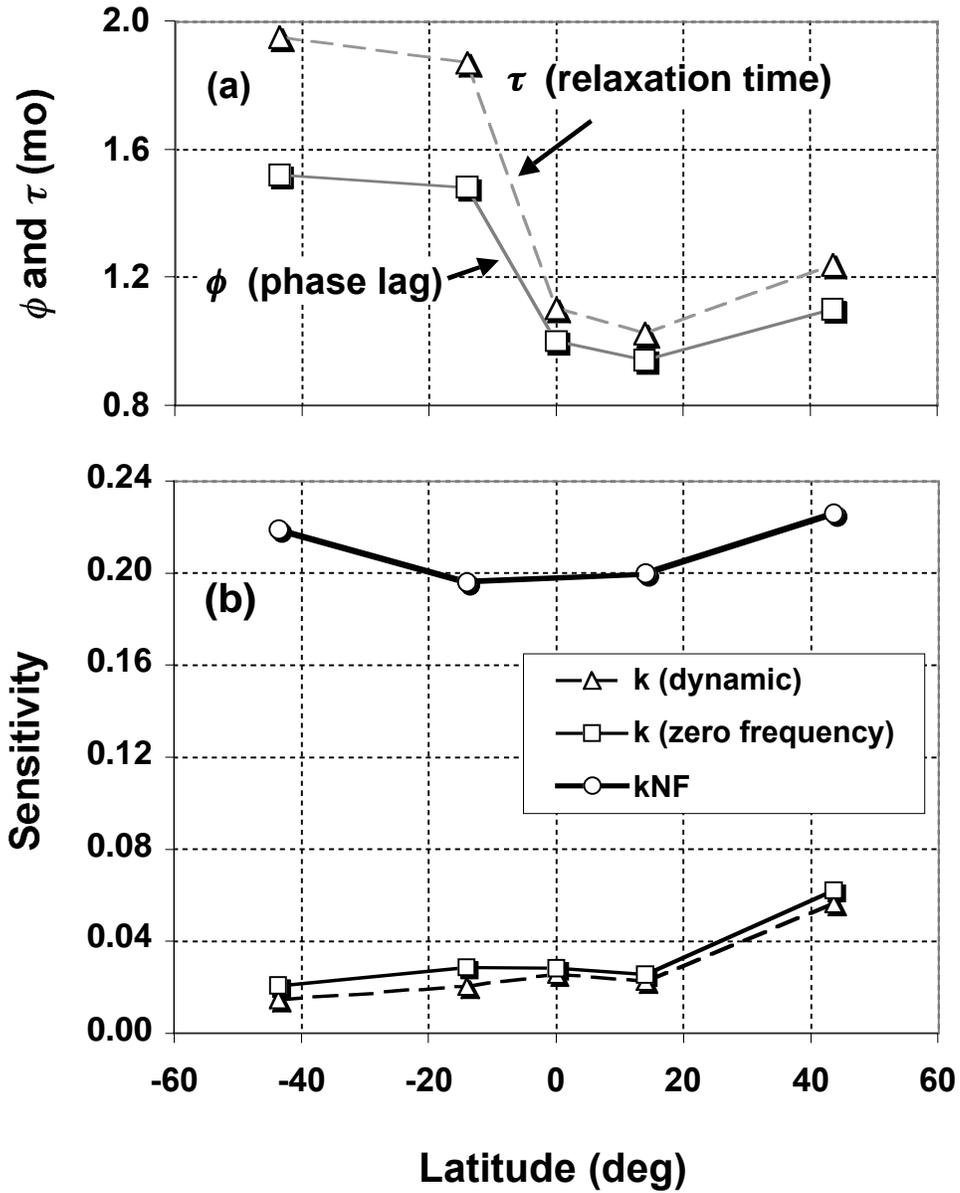